\documentclass[aps,pre,twocolumn,longbibliography,superscriptaddress]{revtex4-1}
\usepackage[latin1]{inputenc}
\usepackage[english]{babel}
\usepackage{bm,graphicx,graphics,amsmath,amssymb,epsfig,color}
\usepackage{dcolumn}

\newcommand{\mA}{\mathcal{A}}

\newcommand{\mD}{\mathcal{D}}

\newcommand{\mK}{\mathcal{K}}

\newcommand{\mM}{\mathcal{M}}

\newcommand{\mZ}{\mathcal{Z}}
\newcommand{\mG}{\mathcal{G}}
\newcommand{\be}{\begin{equation}}
\newcommand{\ee}{\end{equation}}
\newcommand{\bea}{\begin{eqnarray}}
\newcommand{\eea}{\end{eqnarray}}
\newcommand{\bse}{\begin{subequations}}
\newcommand{\ese}{\end{subequations}}
\newcommand{\gcol}[1]{{\color{black} #1}}
\newcommand{\ggcol}[1]{{\color{black} #1}}

\newcommand{\comment}[1]{}
\begin{document}

\title{Symplectic quantization II:\\ dynamics of space-time quantum
  fluctuations and the cosmological constant}

\author{Giacomo Gradenigo} \affiliation{Gran Sasso Science Institute,
  Viale F. Crispi 7, 67100 L'Aquila, Italy}
\email{giacomo.gradenigo@gssi.it}

\begin{abstract}
  The symplectic quantization scheme proposed for matter scalar fields
  in the companion paper~\cite{GL2020} is generalized here to the case
  of space-time quantum fluctuations. That is, we present a new
  formalism to frame the quantum gravity problem. Inspired by the
  \emph{stochastic} quantization approach to gravity,
  \emph{symplectic} quantization considers an explicit dependence of
  the metric tensor $g_{\mu\nu}$ on an additional time variable, named
  \emph{\ggcol{intrinsic} time} at variance with the coordinate time
  of relativity, from which it is different. The physical meaning of
  \ggcol{intrinsic} time, which is truly a parameter and not a
  coordinate, is to label the sequence of $g_{\mu\nu}$ quantum
  fluctuations at a given point of the four-dimensional space-time
  continuum. For this reason symplectic quantization necessarily
  incorporates a new degree of freedom, the derivative
  $\dot{g}_{\mu\nu}$ of the metric field with respect to
  \ggcol{intrinsic} time, corresponding to the conjugated momentum
  $\pi_{\mu\nu}$. Our proposal is to describe the quantum fluctuations
  of gravity by means of a symplectic dynamics generated by a
  generalized action functional $\mA[g_{\mu\nu},\pi_{\mu\nu}] =
  \mK[g_{\mu\nu},\pi_{\mu\nu}] - S[g_{\mu\nu}]$, playing formally the
  role of a Hamilton function, where $S[g_{\mu\nu}]$ is the standard
  Einstein-Hilbert action while $\mK[g_{\mu\nu},\pi_{\mu\nu}]$ is a
  new term including the kinetic degrees of freedom of the field. Such
  an action allows us to define an ensemble for the quantum
  fluctuations of $g_{\mu\nu}$ analogous to the microcanonical one in
  statistical mechanics, with the only difference that in the present
  case one has conservation of the generalized action
  $\mA[g_{\mu\nu},\pi_{\mu\nu}]$ and not of energy. Since the
  Einstein-Hilbert action $S[g_{\mu\nu}]$ plays the role of a
  potential term in the new \emph{pseudo}-Hamiltonian formalism, it
  can fluctuate along the symplectic action-preserving dynamics. These
  fluctuations are the quantum fluctuations of $g_{\mu\nu}$. Finally,
  we show how the standard path-integral approach to gravity can be
  obtained as an approximation of the symplectic quantization
  approach.  By doing so we explain how the integration over the
  conjugated momentum field $\pi_{\mu\nu}$ gives rise to a
  cosmological constant term in the path-integral approach.
  \end{abstract}

	
\maketitle

\section{Introduction}

A general consensus on a quantum gravity theory has still to be found,
although very promising candidates such as string
theory~\cite{P81,GSW87,P98,AGMOO00}, loop quantum gravity~\cite{
  A86,RS90,RS95,R04,T07}, causal dynamic
triangulations~\cite{ADJJ97,AJL04,AGJL12} and non-perturbative
renormalization group approaches~\cite{NR06,N07} are already in the
arena. In our opinion any attempt to correctly frame the study of
quantum gravity must cope in first instance with the problem of
\emph{time} in general relativity.  The \emph{question of time} comes
first. When we talk about \emph{``dynamics''} of the gravitational
field we refer to the possibility for the metric field
$g_{\mu\nu}({\bf p})$ to experience a sequence of changes at a given
point ${\bf p} \in \mM$ of the four-dimensional space-time manifold
$\mM$. The existence of such a sequence of changes is indeed very
natural when thinking, for instance, to any numerical protocol which,
starting from an arbitrary configuration of the field $g_{\mu\nu}({\bf
  p})$, projects it by means of an iterative procedure onto a certain
solution of the Einstein equation, the one corresponding to a given
distribution of matter and energy. In this case the dynamics is the
one of the algorithm iterative procedure and \emph{time} is the number
of iteration steps.  It is evident that this \emph{``time of the
  computer''}, which, as an evolution parameter, closely resembles the
notion of time we have from classical and quantum mechanics, is
totally different from the \emph{coordinate time} of relativity, both
special and general, which is simply a coordinate and not a parameter,
as the name itself suggests. This fact was already well recognized
within the attempt to frame general relativity into the Schr\"odinger
representation of quantum mechanics by means of the Wheeler-DeWitt
equation $\hat{H}\psi=0$~\cite{DeWitt67,DW79}, where $\hat{H}$ is the
general-relativistic Hamiltonian of the gravitational field, $\psi$
the world wave-function and the time derivative term, usually
appearing in the Schr\"odinger equation, is set to zero. This is done
precisely for the reason that in a relativistic context time is just a
coordinate, not an evolution parameter as in quantum mechanics. The
Wheeler DeWitt equation, consistently with Einstein equations, is just
a \emph{constraint} equation, not an evolution equation: it is the
wave equation for the frozen block universe of Einstein, there is no
time flowing.
It is then not by a coincidence that the Wheeler-DeWitt canonical
approach to quantum gravity gave rise to Loop Quantum Gravity
(LQG)~\cite{A86,RS90,RS95,R04,T07}, a theory characterized by the
absence of time at the microscopic level~\cite{R91}. But here we are
not interested in the microscopic degrees of freedom of space-time. We
want to understand which sort of thing could be, physically and not
just in a computer simulation, the \emph{dynamics} of the field
$g_{\mu\nu}$. The answer turns out to be quite simple: as \emph{the
  dynamics} of $g_{\mu\nu}$ we can simply refer to the sequence of its
quantum fluctuations. And as the corresponding \emph{time} parameter
we opt in favour of the \emph{\ggcol{intrinsic} time} introduced~\cite{GL2020}:
the \emph{fifth} variable of symplectic quantization. Consider in fact
the classical Einstein equations for pure gravity:
\begin{align}
  R_{\mu\nu} - \frac{1}{2} g_{\mu\nu} R = 0,
  \label{eq:Einstein}
\end{align}
where $R_{\mu\nu}$ is the Ricci tensor and $R$ is the curvature
scalar. One may think to a \emph{gedanken} experiment where the field
$g_{\mu\nu}$ is set to a very atypical configuration across the whole
space-time manifold $\mM$ and then is let free to relax towards a
solution of the Einstein equations by means of its quantum
fluctuations. When we talk about the dynamics of $g_{\mu\nu}$ we are
of course not talking about a sequence of frames along a time-like
trajectory in the space-time manifold but rather about the sequence of
changes experienced by $g_{\mu\nu}$ due to quantum fluctuations at
each point ${\bf p} \in \mM$ of space-time. The existence of a
dynamics for the gravitational field quantum fluctuations represents a
compelling logical necessity as soon as one accepts the existence of
quantum fluctuations themselves: there must be a process, be it
stochastic or deterministic, which takes the field through their
sequence.
We propose here a generalization to gravity of the symplectic
quantization procedure introduced in~\cite{GL2020} for a scalar field
theory.  The basic idea is the same of~\cite{GL2020}: we try to define
a \emph{pseudo}-microcanonical ensemble built on the conservation of
an appropriately defined generalized action rather than on the
conservation of energy. This generalized action, which formally plays
the role of a Hamilton function, is the generator of a symplectic
dynamics, the dynamics of quantum fluctuations.\\

As done in the companion paper~\cite{GL2020} we introduce symplectic
quantization for the gravitational field by recalling first the
results on the \emph{stochastic} quantization approach to gravity
reported in~\cite{R86}. This is because symplectic quantization stems
from stochastic quantization. We need in fact to acknowledge that the
\emph{``problem of time''} in general relativity has been in a sense
``already solved'' within the stochastic quantization approach in
general~\cite{PW81} and in particular in the application of these
ideas to gravity~\cite{R86}. In fact the true notion of gravitational
field dynamics as the ordered sequence of its quantum fluctuations was
already present in~\cite{R86}.  Unfortunately, at the time the
stochastic quantization approach to gravity was proposed~\cite{R86},
the parameter of the stochastic process controlling the dynamics of
$g_{\mu\nu}$ was just considered a fictitious variable.  On the
contrary, symplectic quantization promotes this fictitious time to be
a true physical entity. \ggcol{Furthermore, symplectic quantization
  allows us to define a functional approach to quantum gravity [see,
    e.g., Eq.~(\ref{eq:micro-prob}) below] which is well-defined
  irrespectively to the scale-invariance properties of the theory, in
  such a way that the \emph{non-renormalizability} problem is
  apparently less compelling in the present context.}\\

We start by discussing first the stochastic quantization approach to
gravity in Sec.~\ref{sec1}. Then in Sec.~\ref{sec2} we introduce
symplectic quantization and in Sec.~\ref{sec3} we show how, within the
symplectic quantization framework, the microcanonical partition
function of a theory without cosmological constant can be mapped onto
the path integral of an Einstein-Hilbert action \emph{with}
cosmological constant $\Lambda$. Sec.~\ref{sec4} is left for a
concluding summary.\\


\section{Stochastic quantization}
\label{sec1}

\gcol{Since the proposal of symplectic quantization has been mainly
  inspired from stochastic quantization, let us start with a
  discussion of the latter, in particular of its application to the
  problem of quantum gravity~\cite{R86}.} The main goal of stochastic
quantization is to introduce a stochastic equation for the metric
field in terms of a fifth parameter $\tau$, named \emph{fictitious}
time in~\cite{R86} and termed by us \emph{\ggcol{intrinsic} time}
in~\cite{GL2020}:
\begin{align}
  \frac{\partial g_{\mu\nu}(x,\tau)}{\partial\tau} = - \int d^4y~G_{\mu\nu,\alpha\beta}(x,y,\tau)~\frac{\delta S[{\bf g}]}{\delta g_{\alpha\beta}(y,\tau) } + \xi_{\mu\nu}(x,\tau),
  \label{eq:Lang}
\end{align}
where the symbol ${\bf g}$ denotes the metric tensor and $S$ is
the Euclidean version of the Einstein-Hilbert action,
\begin{align}
  S[{\bf g}] = \frac{1}{2\kappa} \int d^4x~|g|^{1/2}~R.
  \label{eq:Einstein-Hilbert}
\end{align}
In Eq.~(\ref{eq:Einstein-Hilbert}) $R$ is the Ricci curvature scalar,
the symbol $g$ on the right-hand side denotes the determinant of the
metric tensor, $g=\text{det}(g_{\mu\nu})$, $\kappa = 8 \pi G/c^4$ is the
Einstein gravitational constant and $\xi_{\mu\nu}(x,\tau)$ is a
\emph{white} noise:
\begin{align}
  \langle \xi_{\mu\nu}(x,\tau) \xi_{\alpha\beta}(y,\tau')\rangle =
  \delta(\tau-\tau')~G_{\mu\nu,\alpha\beta}(x,y,\tau).
\end{align}
The symbol $G_{\mu\nu,\alpha\beta}(x,y,\tau)$ denotes the
\ggcol{DeWitt supermetric in four-dimensional space-time, introduced
  in~\cite{DW79} with the purpose to define an invariant measure for
  the path-integral approach to quantum gravity}. Its dependence on $\tau$
comes from the fact that $G_{\mu\nu,\alpha\beta}(x,y,\tau)$ is, as we
are going to see, a function of $g_{\mu\nu}(x,\tau)$. By using the
natural units $\hbar = c = 1$ and choosing the convention according to
which the metric tensor $g_{\mu\nu}$ is dimensionless we have that in
Eq.~(\ref{eq:Einstein-Hilbert}) the Einstein gravitational constant
has the dimensions of a length squared, $[\kappa]=L^2$, while the
Ricci curvature scalar has the dimensions of an inverse length
squared, $[R] = L^{-2}$. The main result of~\cite{R86} is the
perturbative evaluation of metric field correlation amplitude in the
$\tau\rightarrow\infty$ limit:
\begin{align}
& \lim_{\tau\rightarrow\infty} \langle
  g_{\mu\nu}(x,\tau)g_{\alpha\beta}(y,\tau)\rangle = \langle
  g_{\mu\nu}(x)g_{\alpha\beta}(y)\rangle
\label{eq:amplitude}
\end{align}
where the amplitude on the left hand side of Eq.~(\ref{eq:amplitude})
is computed averaging over the time-dependent probability distribution
of the metric field obtained by solving the Fokker-Planck equation
associated to Eq.~(\ref{eq:Lang}). Not surprisingly an exact solution
of this Fokker-Planck equation is out of reach and the results are
only perturbative~\cite{R86}. The problem of non-renormalizability is
also not cured, since $g_{\mu\nu}$ is allowed to fluctuate at all
scales. For gravity the only true advantage of stochastic quantization
is to remove the need of gauge fixing procedures~\cite{PW81}: gauge
orbits are naturally explored by the dynamics, the latter still
yielding the euclidean path-integral measure in the
$\tau\rightarrow\infty$ limit. The good term of comparison to
understand why stochastic quantization removes the problem of gauge
fixing is to consider the overdamped Langevin dynamics of a colloidal
particle confined by a potential $U(x)$:
\begin{align}
\dot{x} = -\frac{\partial U}{\partial x} + \eta(t),
\label{eq:coll-Lang}
\end{align}
where $\eta(t)$ is a white noise of amplitude $\langle \eta^2\rangle
=2\beta^{-1}$. The ``gauge invariance'' hidden in equilibrium
statistical mechanics amounts to the fact that any realization of the
Gaussian noise $\eta(t)$ allows us to sample correctly the equilibrium
distribution $e^{-\beta U(x)}$ in the $t\rightarrow\infty$
limit. Things work in the same manner for the euclidean path-integral
measure of a field theory and stochastic quantization.\\

Let us now spend a couple of words on the bi-tensor density
$G_{\mu\nu,\alpha\beta}(x,y,\tau)$ of Eq.~(\ref{eq:Lang}), where it
guarantees invariance of the equation under the action of
diffeomorphisms in $3+1$ dimensions. In particular, in the general
case we have $G_{\mu\nu,\alpha\beta}(x,y,\tau) \sim
\delta^{(4)}(x,y)$, so that the Langevin equation can be written as:
\begin{align}
  & G^{\mu\nu,\alpha\beta}[{\bf g}(x,\tau)]\dot{g}_{\mu\nu}(x,\tau) = \nonumber \\
  & - \frac{\delta S[{\bf g}]}{\delta g_{\alpha\beta}(y,\tau)} + G^{\mu\nu,\alpha\beta}[{\bf g}(x,\tau)] \xi_{\mu\nu}(x,\tau),
  \label{eq:Lang-2}
\end{align}
where the bitensor density $G^{\mu\nu,\alpha\beta}[{\bf g}(x,\tau)]$
is defined as~\cite{DW79,R86}:
\begin{align}
  & G^{\mu\nu,\alpha\beta}[{\bf g}(x,\tau)] = \nonumber \\
  & \frac{C}{2} |g|^{1/2} \left[ g^{\mu\alpha} g^{\nu\beta} + g^{\mu\beta} g^{\nu\alpha} + \lambda~g^{\mu\nu} g^{\alpha\beta} \right],
  \label{eq:bitensor}
\end{align}
with $\lambda\neq 1/2$ a dimensionless constant and the dependence on
$\tau$ comes clearly by the one of $g_{\mu\nu}(x,\tau)$. The bitensor
density in Eq.~(\ref{eq:bitensor}) represents a field of $d(d+1)/2
\times d(d+1)/2-$dimensional matrices $\hat{G}[{\bf g}(x,\tau)]$ on the
space-time manifold, each matrix having determinant~\cite{DW79}:
\begin{align}
\det \hat{G}[{\bf g}(x,\tau)] = (-1)^{d-1}\left( 1 + \frac{d\lambda}{2}\right)~g^{\frac{1}{4}(4-d)(d+1)}.
\label{eq:detG}
\end{align}
Quite remarkably, the expression of $\det \hat{G}[{\bf g}(x,\tau)]$ is such that it
turns out to be simply a constant \emph{only} in $d=4$:
\begin{align}
\det \hat{G}[{\bf g}(x,\tau)] = - \left( 1 + 2 \lambda \right).
\end{align}
It is the above property of the determinant which guarantees the
\emph{triviality} of the path-integral measure for
gravity~\cite{DW79}. In the framework of stochastic quantization no
dependence on $\tau$ is assumed for the adimensional constant
$\lambda$~\cite{R86}, so that we are going to do the same for the
symplectic quantization approach to be introduced in
Sec.~\ref{sec2}. The importance of chosing a bitensor density
$G^{\mu\nu,\alpha\beta}[{\bf g}(x,\tau)]$ precisely of the form
written in Eq.~(\ref{eq:bitensor}) \ggcol{was noticed already before
  the attempt to apply stochastic quantization to gravity~\cite{R86}
  and was related to the definition of an invariant measure for
  gravitational path-integral~\cite{DW79}}:
\begin{align}
  \mZ_{\text{grav}} = \int e^{i S[{\bf g}]/\hbar}~\mu\left[ {\bf g} \right]~\mD g,
  \label{eq:Zg}
\end{align}
where $\mD g = \prod_x \prod_{\mu\nu} g_{\mu\nu}(x)$ and where
$\mu\left[ {\bf g} \right]$ is a functional measure. If one, in facts,
takes the functional measure $\mu[{\bf g}]$ to be invariant under the action of
diffeomorphisms, i.e., of the form
\begin{align}
  \mu\left[ {\bf g}\right] ~\propto~\prod_{x}\det \hat{G}[{\bf g}(x)],
  \label{eq:mug}
\end{align}
the choice of the operator $\hat{G}[{\bf g}(x)]$ made in
Eq.~(\ref{eq:bitensor}) yields that $\det\hat{G}[{\bf g}(x)]$ is 
constant in $d=4$. Accordingly, the functional measure $\mu[{\bf g}]$
is also constant. The dependence from $\tau$ of ${\bf g}(x,\tau)$ has been
obviously dropped in the partition sum of Eq.~(\ref{eq:Zg}) and in the
expression of the measure, Eq.~(\ref{eq:mug}).  The deep physical
implications of this apparently technical motivation for the choice of
$G^{\mu\nu,\alpha\beta}[{\bf g}(x)]$ will be fully highlighted in
Sec.~\ref{sec3}.\\

From the point of view of dimensional analysis, the requirement of
homogeneity among all elements in Eq.~(\ref{eq:Lang-2}) implies that
$[G^{\mu\nu,\alpha\beta}] = L^{-1}$ in the natural units where
$[\tau] = L$. As a consequence, the constant $C$ in the definition of
$G^{\mu\nu,\alpha\beta}[{\bf g}(x,\tau)]$ carries the dimension of an inverse
length-scale (inverse time-scale):
\begin{align}
[C]  = L^{-1}.
\end{align}

Only two comments are now in order before moving to the discussion of
symplectic quantization. In our opinion the main breakthrough of
stochastic quantization was the proposition of the Langevin equation
in Eq.~(\ref{eq:Lang}). In particular, stochastic quantization gave a
clear indication of a plausible solution for the \emph{``problem of
  time''} in general relativity. Quite unfortunately, since the
additional time variable needed to this purpose was not regarded as a
physical entity, but merely as a technical device (disappearing at the
end of the calculation), this insightful approach was quite
underrated. From the conceptual point of view, the main problems of
stochastic quantization were related to the properties of
$\xi_{\mu\nu}(x,\tau)$, only fixed by the requirement of overall
self-consistency for the theory and apparently lacking any physical
interpretation. In fact, quite obviously, the presence of
$\xi_{\mu\nu}(x,\tau)$ in Eq.~(\ref{eq:Lang}) does not represent the
action of a thermostat. In conclusion, any possible deep conceptual
implication of the stochastic quantization approach to gravity was
ruled out for the reason that the \emph{fifth} time variable was
regarded just a fictitious one, a fact related to the lack of a
physical interpretation for $\xi_{\mu\nu}(x,\tau)$. On the technical
side the theory was working, but not remarkably better than other
proposals. In particular the worst plague of quantum gravity, the
non-renormalizability problem, was still there. There have been some
very interesting recent progresses on the stochastic quantization
approach to gravity, where higher order derivatives with respect to
the fictitious time were added to the Langevin equation for
$g_{\mu\nu}$ and where it is clearly alluded to some physical
relevance of the fictitious time~\cite{BW20}. Nevethertheless, as long
as the presence of noise is a cornerstone of the approach
proposed~\cite{BW20}, the intepretative problems of the fictitious
time remain. We will show in the next section that both conceptual and
technical problems related to noise completely disappear in the
context of symplectic quantization, which is a deterministic
process.
\ggcol{Furthermore, since the value of the generalized action is fixed
  and the cornerstone of the approach is an action preserving
  symplectic dynamics, the non-renormalizability problem of gravity is
  possibly less severe in this context than in the Euclidean
  path-integral approach. The reason is that alongside with symplectic
  quantization one introduces a sort of microcanonical measure which
  is well defined irrespectively to whether the theory is or not
  renormalizable.}

\section{Symplectic quantization}
\label{sec2}

The key idea of symplectic quantization is to replace the stochastic
dynamics of Eq.~(\ref{eq:Lang}) with a deterministic one. This choice,
which is motivated and explained in great detail in~\cite{GL2020},
comes from the following argument. The Langevin equation of stochastic
quantization is a fictitious stochastic dynamics which allows one to
sample, for asymptotically long times, field configurations with the
Boltzmann-like weight appearing in the euclidean version of the
path-integral for gravity. In~\cite{GL2020} it has been shown that the
``canonical'' ensemble which is in that way implicitly defined by
stochastic quantization can, and sometimes has to, be replaced with a
\emph{fixed-action ``microcanonical'' ensemble}, for which an
appropriate generalized action has to be drawn~\cite{GL2020}. This
generalized action is obtained by adding the kinetic terms containing
the derivative of the field with respect to \ggcol{intrinsic} time,
usually absent in any standard field theoretic action $S$, which
usually contains derivatives only with respect to space-time
coordinates. \ggcol{Soon after the first appearance of the present
  work as a preprint on public repositories we knew~\cite{Caracciolo}
  of a previous attempt to build a generalized action for
  quantum fields containing momenta with respect to a fifth time
  variable: the functional approach to field theory proposed by De
  Alfaro, Fubini and Furlan in 1983~\cite{DFF83}. The additional
  kinetic term therein was build according to a logic similar to the
  one followed here, but making different choices on the technical
  level (vierbeins $V^a_\mu$ were used instead of $g_{\mu\nu}$ and the
  DeWitt supermetric was not considered). Then, most importantly,
  in~\cite{DFF83} the symplectic dynamics and the microcanonical
  ensemble introduced in~\cite{GL2020} and generalized here to the
  case of gravity were never considered nor alluded and the fifth time
  variable was \emph{not} considered a physical one. Coming back to
  the present proposal, we have that for} gravity the~\emph{kinetic}
term corresponding to the one of Eq.~(20) in~\cite{GL2020} has to be
an object quadratic in $\dot{g}_{\mu\nu}$, where the dot now indicates
derivative with respect to \ggcol{intrinsic} time $\tau$. At variance
with~\cite{GL2020}, where the only constraint on the new kinetic term
was to be a Lorentz scalar, in the case of gravity we need it to be a
scalar with respect to the action of diffeomorphisms. As the key
ingredient to define a \emph{kinetic} term fulfilling this requirement
we used the DeWitt supermetric $G^{\mu\nu,\alpha\beta}[{\bf
    g}(x,\tau)]$~\cite{DW79}. Taking inspirantion from the analysis
of~\cite{GL2020} we propose a generalized action for the gravitational
field where $S[{\bf g}]$ is treated as a potential term:
\begin{align}
  \mA[{\bf g},\dot{{\bf g}}] = \mK[{\bf g},\dot{{\bf g}}] - S[{\bf g}],
  \label{eq:pseudo-H}
\end{align}
with
\begin{align}
  & \mK[{\bf g},\dot{{\bf g}}] = \nonumber \\
  & \frac{1}{2\kappa_{g}}
  \int d^4x~|g|^{1/2}~\dot{g}_{\mu\nu}(x,\tau)~\mathcal{G}^{\mu\nu,\alpha\beta}[{\bf g}(x,\tau)]\dot{g}_{\alpha\beta}(x,\tau),
  \label{eq:kin-lagrange}
\end{align}
where we have introduced the dimensionless supermetric tensor
$\mathcal{G}^{\mu\nu,\alpha\beta}[{\bf
    g}(x,\tau)]=G^{\mu\nu,\alpha\beta}[{\bf g}(x,\tau)]/C$. Due to the
properties of the supermetric the expression in
Eq.~(\ref{eq:kin-lagrange}) clearly behaves as a scalar under the
action of diffeomorphisms. $S[{\bf g}]$ is the Einstein-Hilbert
action. In particular $S[{\bf g}]$ is the \emph{original}
Einstein-Hilbert action for a Lorentzian space-time, not the Euclidean
version of the action. In fact, as explained in full detail
in~\cite{GL2020}, symplectic quantization allows one to define the
probability of configurations directly in a space-time with
Minkowskian signature, with no need to consider a Wick rotation to the
Euclidean theory. For dimensional reasons we have introduced the
constant $\hat{\kappa}_g$ with dimensions $[\hat{\kappa}_g]=L$. If we
regard the generalized action $\mA[{\bf g},\dot{{\bf g}}]$ as
\emph{pseudo}-Hamiltonian which generates the evolution in
\ggcol{intrinsic} time $\tau$, in Eq.~(\ref{eq:pseudo-H}) the
Einstein-Hilbert action $S[{\bf g}] $ clearly plays in the role of a
potential term, since it contains no \ggcol{intrinsic} time derivative
terms. In order to dissipate any possible doubt, let us stress that
\ggcol{intrinsic} time $\tau$ is a parameter, a sort of
\emph{internal} degree of freedom of the metric field, so that the
kinetic term $\mK[{\bf g},\dot{{\bf g}}]$ has to be a scalar under the
action of diffeomorphisms in $3+1$ dimensions, \emph{not} in $3+2$
dimensions. We are not adding extra-dimensions to space-time. In order
to introduce Hamilton equations it is convenient to rewrite $\mK[{\bf
    g},\dot{{\bf g}}]$ in terms of the conjugated momentum
\begin{align}
\pi^{\mu\nu}(x) = \frac{1}{\kappa_g}\mathcal{G}^{\mu\nu,\alpha\beta}[{\bf g}(x,\tau)]~\dot{g}_{\alpha\beta}(x,\tau).
\label{eq:pi}
\end{align}
We thus have
\begin{align}
  & \mK[{\bf g},\boldsymbol{\pi}] = \nonumber \\
  & \frac{\kappa_g}{2} \int d^4x~|g|^{1/2}~\pi^{\mu\nu}(x,\tau)~\mathcal{G}_{\mu\nu,\alpha\beta}[{\bf g}(x,\tau)]~\pi^{\alpha\beta}(x,\tau),
\label{eq:kinetic}
\end{align}
or, what is equivalent,
\begin{align}
  & \mK[{\bf g},\boldsymbol{\pi}] = \nonumber \\
  &\frac{\kappa_g}{2} \int d^4x~|g|^{1/2}~\pi_{\mu\nu}(x,\tau)~\mathcal{G}^{\mu\nu,\alpha\beta}[{\bf g}(x,\tau)]~\pi_{\alpha\beta}(x,\tau),
\label{eq:kinetic-2}
\end{align}
where, to get Eq.~(\ref{eq:kinetic}) from Eq.~(\ref{eq:pi}) and
Eq.~(\ref{eq:kin-lagrange}), we have used the identity
\ggcol{$\mathcal{G}_{\mu\nu,\alpha\beta} \mathcal{G}^{\mu\nu,\alpha\beta}=d(d+1)/2$} and
to get Eq.~(\ref{eq:kinetic-2}) we have raised/lowered indices with
the help of the metric tensor, taking always advantage of the identity
\ggcol{$g_{\mu\nu}g^{\mu\nu}=d$}:
\begin{align}
  \mathcal{G}_{\mu'\nu',\alpha'\beta'} & = \mathcal{G}^{\mu\nu,\alpha\beta} g_{\mu\mu'}g_{\nu\nu'}g_{\alpha\alpha'}g_{\beta\beta'} \nonumber \\
  \pi_{\mu'\nu'} & = \pi^{\mu\nu} g_{\mu\mu'}g_{\nu\nu'}.
\end{align}
\ggcol{In Eq.~(\ref{eq:kinetic}) constants coming from the contraction
  of tensor indices have been adsorbed into the definition of
  $\kappa_g$}. Since the supermetric is now dimensionless,
$[\mathcal{G}^{\mu\nu,\alpha\beta}] = 0$, and in natural units intrinsic
time has dimensions of length, we have that the intrinsic time derivative
of the metric field has dimensions
\begin{align}
  \dot{g}_{\mu\nu} =  L^{-1},
\end{align}
while its conjugated momentum $[\pi^{\mu\nu}]=[\mathcal{G}^{\mu\nu,\alpha\beta}\dot{g}_{\alpha\beta}/\kappa_g]$ has dimensions
\begin{align}
[\pi^{\mu\nu}]=L^{-3}
\end{align}
Let us notice that, despite having the same physical dimension, the
Einstein gravitational constant $\kappa$ and the new dimensional
constant $\kappa_g$ appearing in $\mK[{\bf g},\dot{{\bf g}}]$ do not
need to be the same. And in fact we will see they are different
constants. We also assume, consistently with the stochastic
quantization approach proposed in~\cite{R86}, that the adimensional
constant $\lambda$ appearing in the definition of
$G^{\mu\nu,\alpha\beta}[{\bf g}(x,\tau)]$ is a constant with respect
to \ggcol{intrinsic} time $\tau$.

A close look at Eq.~(\ref{eq:kinetic-2}) reveals that the new
generalized action, which we may also call the
\emph{pseudo}-Hamiltonian of symplectic quantization, is non-separable
because the \ggcol{DeWitt supermetric} depends on the metric field
${\bf g}(x,\tau)$. This \emph{non-separability} of the
pseudo-Hamiltonian is the most remarkable difference between the
symplectic quantization of gravity and that of a prototypical matter
field discussed in~\cite{GL2020}. We will see in Sec.~\ref{sec3} that
this fact has very interesting consequences. For the moment let us
write down the Hamilton equations which, according to the symplectic
quantization scenario, govern the dynamics of the gravitational field
quantum fluctuations:
\begin{align}
  \dot{g}^{\mu\nu}(x,\tau) &= \frac{\delta \mA[{\bf g},\boldsymbol{\pi}]}{\delta \pi_{\mu\nu}(x,\tau)}  \nonumber \\
  \dot{\pi}^{\mu\nu}(x,\tau) &= - \frac{\delta \mA[{\bf g},\boldsymbol{\pi}]}{\delta g_{\mu\nu}(x,\tau)}  \nonumber \\
  \label{eq:Hamilton}
\end{align}  
The quantity $\mA[{\bf g},\boldsymbol{\pi}]$ is fixed by the
choice of initial conditions, due to the symplectic nature of
dynamics. This means that the quantum fluctuations of the
gravitational field, which are nothing but the fluctuations of the
\emph{``potential''} term $S[{\bf g}]$ driven by the action-preserving
dynamics of Eq.~(\ref{eq:Hamilton}), are only those compatible with
the conservation of $\mA[{\bf g},\boldsymbol{\pi}]$. 

If then, rather than following the dynamics of the metric field
fluctuations, one wants to sum other them by assuming the ``bona
fide'' ergodicity of the dynamics in Eq.~(\ref{eq:Hamilton}), the
appropriate statistical ensemble must be chosen. As it has been
already outlined in the companion paper~\cite{GL2020}, for a
relativistic field theory the most appropriate choice is that of a
microcanonical-type of ensemble, defined with respect to the
generalized action $\mA[{\bf g},\boldsymbol{\pi}]$, rather than a
canonical-type one, which would correspond to the standard
path-integral approach. As we did for the scalar field theory
in~\cite{GL2020}, let us point out the main and only assumption at the
basis of the \emph{microcanonical ensemble} for the quantum
fluctuations of space-time:\\

\emph{``All configurations of the tensorial fields ${\bf g}(x)$ and
  $\boldsymbol{\pi}(x)$ which correspond to the same value of the
  generalized action $\mA[{\bf g},\boldsymbol{\pi}]$ are realized with
  identical probability.''}\\

This is the microcanonical postulate for the metric field in the
general relativistic context. Clearly this generalized microcanonical
ensemble has nothing to do with temperature or thermal fluctuations,
since the quantity which is conserved is action, not energy. For this
ensemble the partition function reads as:
\begin{equation}
\Omega(A) = \int \mD g~\mD \pi~\delta\left( A - \mA[{\bf g},\boldsymbol{\pi}] \right),
\label{eq:micro-gravity}
\end{equation}
where $\mD g$ can be simply taken as $\mD g = \prod_x \prod_{\mu\nu}
g_{\mu\nu}(x)$, consistently with the choice of the functional measure
$\mu[{\bf g}]$ shown in Eq.~(\ref{eq:mug}), while $\mD \pi$ is:
\begin{align}
  \mD \pi &= \prod_x \prod_{\mu\nu} \left[ \sqrt{\kappa_g}
    d\pi_{\mu\nu}(x) \right] = \kappa_g^{V/2} ~\prod_x
  \prod_{\mu\nu} d\pi_{\mu\nu}(x),
  \label{eq:dpi}
\end{align}
where the dimensional constant $\kappa_g$ has been added as a factor
for each infinitesimal volume element of functional integration in
order to keep the partition function dimensionless, while the symbol
$V$ in $\kappa_g^{V/2}$ denotes the invariant integration volume.\\

\ggcol{Let us stress that the partition function in
  Eq.~(\ref{eq:micro-gravity}) is well defined irrespectively to the
  renormalizability of the theory. The expression may look formal, but
  it is not. As already outlined in~\cite{GL2020}, the advantage of an
  expression like the one in Eq.~(\ref{eq:micro-gravity}) is to be a
  functional integral with a clear probabilistic interpretation, at
  variance with the Feynman path integral. Then, similarly to the path
  integral, one has the problem of functional integration, but nothing
  more. In particular, while the theoretical justification of path
  integrals in quantum field theory comes only \emph{``\`a posteriori''}
  as long as they allow to compute finite amplitues, and to this
  purpose renormalizability is crucial in order for a theory to be
  well defined, this is not the case for the partition function in
  Eq.~(\ref{eq:micro-gravity}), which comes from first principles and
  has a clear physical interpretation on its own. Within symplectic
  quantization the fundamental level of description of the systems is
  assumed to be, as in classical statistical mechanics, the
  microscopic symplectic dynamics. Then, under the assumption of
  ergodicity for the latter, one can write a partition function such
  as the one in Eq.~(\ref{eq:micro-gravity}). Quantum fluctuations are
  controlled in first place by the dynamics of
  Eq.~(\ref{eq:Hamilton}), from which the microcanonical partition
  function in Eq.~(\ref{eq:micro-gravity}) follows. There might be
  \emph{``pathological''} configurations which satisfy the
  fixed-action constraint by putting a very large, but compensating,
  amount of action respectively in the ``kinetic'', $\mK[{\bf
      g},\dot{{\bf g}}]$, and in the ``potential'', $S[{\bf g}]$,
  terms of the generalized action. But this is not a problem, since
  within this approach the physical consistency of the theory does not
  rely on renormalizability. From such a perspective, where the
  scale-invariance property of the theory is not compelling as usual,
  the problem of functional integration can be easily solved by
  putting the theory on a lattice (recipes for this have been devised
  in the past~\cite{CP89}) and by sampling configurations with
  probability
  \begin{align}
    P_A({\bf g},\boldsymbol{\pi}) =
    \frac{1}{\Omega(A)}\delta\left(  A - \mA[{\bf g},\boldsymbol{\pi}] \right).
    \label{eq:micro-prob}
  \end{align}

  The key point is that the probability density in
  Eq.~(\ref{eq:micro-prob}) comes directly from the original theory
  with no need of Wick-rotating to Eucliden path-integrals.  The
  rotation from real to immaginary time has in fact no physical
  motivation other than allowing us to work with a well-defined
  probability measure. Such a measure allows for non-perturbative
  approaches, for instance to cast strong-coupling expansions on the
  lattice for non-abelian gauge theories as high-temperature
  expansions in statistical mechanics~\cite{MM94}. But the whole
  justification of working with Euclidean field theory comes only
  \emph{``\`a posteriori''}: it is thus very appealing to have a
  functional approach to field theory which already in real time is
  well defined from the point of view of probability. Moreover, recent
  results pertaining precisely the stochastic quantization approach to
  gravity~\cite{BW20}, showed that in certain circumstances (e.g.,
  during inflation) it is not even possible to rotate the Euclidean
  theory back to the original quantum field theory. Therefore, in
  particular for gravity, a formalism allowing us to write functional
  integrals well defined probabilistically already in Minkowski space
  seems particularly useful.}\\

\gcol{\ggcol{The problem of gravity non-renormalizability becomes more
    compelling when we try to change the statistical ensemble of
    quantum fluctuations, that is, when we try to transform the
    microcanonical type of partition sum in
    Eq.~(\ref{eq:micro-gravity}) into a canonical one. For the same
    reason explained in~\cite{GL2020}, namely the non-definiteness of
    a relativistic action sign, the integral transformation which
    allows us to change ensemble must be a Fourier transform:

  \begin{align}
    \mZ_{\text{grav}} = \int dA~e^{-i A/\hbar}~\Omega(A) = \int \mD\pi\mD g~
    e^{-i\hbar \mA[{\bf g},\boldsymbol{\pi}]}.
    \label{eq:first-canonical}
  \end{align}

  For an expression like the one in Eq.~(\ref{eq:first-canonical}) we
  are back to the old problem: it makes sense as long as the
  theory it is renormalizable. Nevertheless now there are two
  intermediate steps which could be forbidden for physical reasons:
  first, the assuption that an equilibrium measure, like the one in
  Eq.~(\ref{eq:micro-gravity}), really exists, assumption which is based on
  the ergodicity of dynamics in Eq.~(\ref{eq:Hamilton}); second, the
  assumption that the ensemble with a \emph{hard} constraint on the
  action, Eq.~(\ref{eq:micro-gravity}), and the ``canonical'' ensemble
  of Eq.~(\ref{eq:first-canonical}), characterized by a \emph{soft}
  constraint, are equivalent. If we believe that in most cases of
  interest ergodicy holds, we are then left with the hypothesis that
  for gravity the two ensembles just mentioned are not
  equivalent. Something not at all suprising, if true, since the lack
  of equivalence of statistical ensembles is not uncommon in
  statistical mechanics. Moreovere, precisely the gravitational
  potential in the non-relativistic regime is a well-known case were
  inequivalence of statistical ensembles for thermal fluctuations
  takes place~\cite{CDR09}. In particular, the non equivalence of
  thermal ensembles for gravitating systems comes from the
  \emph{long-range} nature of the gravitational
  interaction~\cite{CDR09}, which makes energy non-extensive.

  In summary, while it is standard knowledge that different
  statistical ensembles are not equivalent for the large-scale thermal
  fluctuations of gravity~\cite{CDR09}, in the present work we also
  put forward the hypothesis that a similar ``non-equivalence'' holds
  even for the statistical ensembles of quantum fluctuations at high
  energy. In this perspective, as soon as consensus will be gathered
  around one of the high-energy regulators proposed for a quantum
  theory of gravity (e.g., loop quantum
  gravity~\cite{R04,T07}), this finding will also have an impact on
  the problem of statistical ensembles inequivalence for the quantum
  fluctuations of gravity. But, since a general agreement on such a UV
  regulator has not been reached so far, we rely here on the most
  \emph{agnostic}, at present, hypothesis: the
  \emph{pseudo}-microcanonical ensemble of
  Eq.~(\ref{eq:micro-gravity}) and the path-integral of
  Eq.~(\ref{eq:first-canonical}) are not equivalent. In what follows
  the path-integral approach to gravity will be thus regarded just as
  a low-energy approximation of the theory.}\\

In~\cite{GL2020} we have shown that the \emph{pseudo}-microcanonical
partition function for a relativistic scalar field reduced to the
Feynman path-integral of the corresponding field theory, after Fourier
transforming with respect to the action and integrating over
momenta. In what follows we are going to do the same for the
\emph{pseudo}-microcanonical partition function of
Eq.~(\ref{eq:micro-gravity}).}

\section{Cosmological constant}
\label{sec3}

We expect that the main goal of symplectic quantization will be
represented by the possibilities it opens up to describe in a
consistent manner the \emph{dynamics} of the gravitational field in
all situations where this can be relevant. For instance, we have in
mind the non-equilibrium aspects of the quantum fluctuations
relaxational dynamics in inflationary
cosmology~\cite{MMOL94,GLMM94,PRT20,PVWA21,DLRG2021}. We also expect
that symplectic quantization will allow us to give a simple and
unequivocal definition in the framework of quantum cosmology of
concepts typical of statistical systems such as \emph{non-equilibrium
  dynamics} and \emph{irreversibility}. \ggcol{For example, we expect
  symplectic quantization to be the good formalism to go beyond the
  Einstein's frozen-block universe scenario, i.e., the good formalism
  to represent an \emph{evolutionary} dynamics of the
  universe~\cite{S92}.} But right now we leave these speculations as a
matter for future investigations and we present an analysis more
limited in scope. We want to show how the new kinetic degrees of
freedom introduced in the context of symplectic quantization are
directly related to the appearance of a cosmological constant term in
Einstein equations passing via the path-integral approach to
gravity. In particular we want to show that the symplectic
quantization of pure gravity \emph{without} a cosmological constant
term produces, by simply integrating over the conjugated momenta
$\pi_{\mu\nu}(x)$, a theory of gravity \emph{with} a cosmological
constant $\Lambda$. This derivation of $\Lambda$ is therefore still
quantum in nature but is \emph{different} from the usual one, where
$\Lambda$ is interpreted as the vacuum energy of matter fields: here
$\Lambda$ is solely related to intrinsic properties of pure gravity,
in particular to the fact that $g_{\mu\nu}$ has its own
dynamics. \ggcol{Let us stress that the procedure outline here, though
  provinding a new physical interpretation and derivation of the
  cosmological constant, is in first instance motivated by the
  necessity to relate the new kinetic degrees of freedom of the field,
  $\pi_{\mu\nu}(x)$, so far unobserved, to known physics.}\\

The procedure which allows us to relate the momenta $\pi_{\mu\nu}(x)$
to the cosmological constant requires to transform the partition
function in Eq.~(\ref{eq:micro-gravity}) from the hard-constraint
ensemble, where the value of $\mA[{\bf g},\boldsymbol{\pi}]$ is fixed
with a Dirac delta, to the ensemble where the constraint is soft, for
instance by means of an integral (Fourier) transform:
\begin{align}
  \mZ_{\text{grav}}(z) &= \int d A~e^{-i z A}~\Omega(A) = \int
  \mD g~\mD \pi ~e^{-i z \mA[{\bf g},\boldsymbol{\pi}]} = \nonumber \\
  & = \int \mD g~\mZ_{\text{kin}}[z,{\bf g}]~e^{i z S[{\bf g}]},
  \label{eq:path-integral}
\end{align}
where we have defined
\begin{align}
  \mZ_{\text{kin}}[z,{\bf g}] = \int \mD \pi~e^{-i z\mK[{\bf g},\boldsymbol{\pi}]}.
  \label{eq:Zg-kin}
\end{align}
The integration over momenta is technically quite easy, since the
pseudo-Hamiltonian $\mA[{\bf g},\boldsymbol{\pi}]$ depends
quadratically on them, but it is at the same time non-trivial due to
the non-separable nature of $\mA[{\bf g},\boldsymbol{\pi}]$. This
non-separability comes from the dependence of the kinetic term
$\mK[{\bf g},\boldsymbol{\pi}]$ on the metric itself, a characteristic
which is typical of the symplectic quantization of gravity and not of
ordinary matter fields. By choosing $z=1$ (in natural units, which
means $z = \hbar^{-1}$), one can write:
\begin{align}
  & \mZ_{\text{kin}}[1,{\bf g}] =\nonumber \\ & = \int \mD
  \pi(x)~e^{-i\frac{\kappa_g}{2} \int
    d^4x~|g|^{1/2}~\pi_{\mu\nu}(x)~\mG^{\mu\nu,\alpha\beta}[{\bf g}(x)]~\pi_{\alpha\beta}(x)}
  \nonumber \\ & = \exp\left( -\frac{i}{2 \kappa_g^2}\int
  d^4x~|g|^{1/2}\log[\text{det}~\hat{\mG}[{\bf g}(x)]] \right).
  \label{eq:main}
\end{align}
Intermediate steps of the not difficult calculation yielding the right
hand side of Eq.~(\ref{eq:main}) are reported in the Appendix.\\

We need now to recall a quite remarkable coincidence: {\it only in
  four space-time dimensions} it happens that the determinant of the
bi-tensor $\hat{\mG}[{\bf g}(x)]$ is constant across the whole
space-time manifold~\cite{R86,DW79}:
\begin{align}
\det\hat{\mG}[{\bf g}(x)] = -1-2\lambda. 
\label{eq:detG}
\end{align}
In~\cite{R86} it is proposed the value $\lambda=-1$ for the
adimensional constant in order to have physical consistency of the
stochastic quantization procedure. In the case of symplectic
quantization we find convenient to ask for a small deviation from this
value, i.e., we propose
\begin{align}
  \lambda = -\left(1 + \frac{\varepsilon_\Lambda^2}{2}\right),
  \label{eq:lambda}
\end{align}
where we take $\varepsilon_\Lambda$, which is the only free parameter
of the new theory, as a small positive number,
$\varepsilon_\Lambda<1$. The choice of the subscript $\Lambda$ in
$\varepsilon_\Lambda$ will be immediately clear.\\

{\it We have achieved an interesting result: the integration over
  the kinetic degrees of freedom of the metric field represents a
  possible way to derive the existence of a cosmological term in
  Einstein equations.}\\

In fact defining
\begin{align}
  \ggcol{2}\Lambda = \frac{\kappa}{\kappa_g^2}\log \text{det}~\hat{\mG}[{\bf g}(x)]
\end{align}
and plugging into Eq.~(\ref{eq:detG}) the definition of $\lambda$
given in Eq.~(\ref{eq:lambda}) we get
\begin{align}
  \ggcol{2}\Lambda = \frac{\kappa}{\kappa_g^2}\log[1+\varepsilon_{\Lambda}^2]
  \simeq \varepsilon^2_{\Lambda} \frac{\kappa}{\kappa_g^2}.
  \label{eq:L-formula}
\end{align}  
The choice made for the parameter $\lambda$ in Eq.~(\ref{eq:lambda})
is therefore consistent with a positive cosmological constant,
$\Lambda > 0$, and we can finally recover the path-integral for the
Einstein-Hilbert action of pure gravity \emph{with} a cosmological
constant term:
\begin{align}
\mZ_{\text{grav}} &= \int \mD g(x)~\mD \pi(x)~e^{-i \mA[{\bf g},\boldsymbol{\pi}]} = \nonumber \\
& = \int \mD g(x) \exp\left( i \frac{1}{2\kappa}\int d^4x~|g|^{1/2} \left[ R - \ggcol{2} \Lambda \right] \right).
\label{eq:cosmological}
\end{align}
The derivation of Eq.~(\ref{eq:cosmological}) from the
\emph{pseudo}-microcanonical partition function in
Eq.~(\ref{eq:micro-gravity}) is the main result of this paper: the
integration over the kinetic degrees of freedom of the field
$\boldsymbol{\pi}(x)$ give rise to a cosmological constant term.\\


We want to recall at this point that in order to define the symplectic
quantization ensemble for the quantum fluctuations of gravity we have
introduced only one new physical constant, $\kappa_g$, the one
appearing in the definition of $\mK[{\bf g},\boldsymbol{\pi}]$ in
Eq.~(\ref{eq:kinetic}). The constant $\lambda$ appearing in the
definition of the DeWitt \ggcol{supermetric}
$\mG^{\mu\nu,\alpha\beta}[{\bf g}(x)]$ is in fact just a dimensionless
number. Now, by reverse engeneering the relation between constants in
Eq.~(\ref{eq:L-formula}), we can see that $\kappa_g$ is not really a
new object but can be written as a combination of known constants:
\begin{align}
  \kappa_g \simeq \sqrt{\frac{\kappa}{\ggcol{2}\Lambda}}.
  \label{eq:kappag-def}
\end{align}
That is, rather than saying that the dynamics of $g_{\mu\nu}(x)$,
which is controlled by $\kappa_g$, give rise to a cosmological
constant term, we can see the relation between constants from the
point of view of Eq.~(\ref{eq:kappag-def}): it is $\Lambda$ that
determines the typical scale for the symplectic dynamics of
${\bf g}(x,\tau)$. In order to make this clear one can rewrite the kinetic
term $\mK[{\bf g},\boldsymbol{\pi}]$, taking advantage of
Eq.~(\ref{eq:L-formula}), as:
\begin{align}
  & \mK[{\bf g},\boldsymbol{\pi}] = \nonumber \\
  & \frac{1}{2\varepsilon_\Lambda}\sqrt{\frac{\ggcol{2}\Lambda}{\kappa}}
  \int d^4x~|g|^{1/2}~\dot{g}_{\mu\nu}(x,\tau)~\mG^{\mu\nu,\alpha\beta}[{\bf g}(x)]~\dot{g}_{\alpha\beta}(x,\tau)
  \label{eq:kinetic-new}
\end{align}  
It is easy to see that the expression in Eq.~(\ref{eq:kinetic-new}) is
dimensionless, as it has to be, since $[\varepsilon_\Lambda] = [\mG] = [g]=0 $,
$[\Lambda]=L^{-2}$, $[\kappa]=L^2$ and $[\dot{g}_{\mu\nu}]=L^{-1}$.\\

In conclusion we have shown how to obtain from the microcanonical
partition function of symplectic quantization,
Eq.~(\ref{eq:micro-gravity}), the standard Feynman path-integral for
gravity~\cite{SH79}, showing that the cosmological constant term in
Einstein equations is a fingerprint of the existence of \emph{kinetic}
degrees of freedom for $g_{\mu\nu}$. The theory proposed is therefore
predictive: it describes a new phenomenon, the existence of an
intrinsic dynamics for the field ${\bf g}(x,\tau)$, and relates a
feature of this new phenomenon, i.e., the characteristic time-scale of
the gravitational field quantum fluctuations dynamics, to known
physical constants: the Einstein gravitational constant $\kappa$ and
the cosmological constant $\Lambda$, which has to be taken as an input
from observational data. In the new theory there is only one
dimensionless parameter let free to be fixed, the constant
$\varepsilon_\Lambda$ appearing in the definition of the DeWitt
supermetric~$\mG^{\mu\nu,\alpha\beta}[{\bf g}(x)]$.\\

\section{Conclusions}
\label{sec4}

Let us summarize here the main logical steps explaining how symplectic
quantization, first proposed for a scalar field in~\cite{GL2020},
works for gravity. The most relevant novelty of the approch is the
claim that fields, in particular the metric field ${\bf g}(x)$ in the
present case, depend on an additional variable $\tau$ with physical
dimensions of time, i.e. we have ${\bf g}(x,\tau)$.  This variable,
here referred to as \emph{intrinsic time} $\tau$, is not an additional
coordinate of the space-time continuum: $\tau$ is the parameter which
controls the sequence of quantum fluctuations at each point of the
space-time manifold and is different from the coordinate time $x^0=
ct$. \emph{Intrinsic} time $\tau$, which indeed already appeared as a
fictitious variable in the stochastic quantization
approach~\cite{PW81,R86}, it is raised here to its full dignity of
physical time. In the framework of symplectic quantization it is
therefore possible to consistently define a \emph{dynamics} for the
quantum fluctuations of ${\bf g}(x,\tau)$. This dynamics, which is
deterministic, is generated by a generalized action of the kind
$\mA[{\bf g},\boldsymbol{\pi}] = \mK[{\bf g},\boldsymbol{\pi}] -
S[{\bf g}]$. While $S[{\bf g}]$, which can be taken as any general
relativistic action for the gravitation field (e.g., Einstein-Hilbert,
but the approach works for modified gravity theories as
well~\cite{CFPS12}), plays the role of a \emph{potential} term, we
have that $\mK[{\bf g},\boldsymbol{\pi}]$ is a \emph{kinetic} term,
controlling the rate of change of ${\bf g}(x,\tau)$ with respect to
\ggcol{intrinsic} time. The term $\mK[{\bf g},\boldsymbol{\pi}]$
appears for the first time in the present paper, up to our
knowledge. The knowledge of $\mA[{\bf g},\boldsymbol{\pi}]$ allows
then us to write down the symplectic dynamics in \ggcol{intrinsic}
time generated by the Hamilton equations in~(\ref{eq:Hamilton}). The
fluctuations of the term $S[{\bf g}]$ along the action-preserving
dynamics represent the quantum fluctuations of the gravitational
field. By then assuming a sort of ergodicity for this symplectic
dynamics, a fact which is far from trivial and might not be true in
some extreme conditions (e.g., close to cosmological singularities and
black holes), it is then possible to define a
\emph{pseudo}-microcanonical ensemble based on the hypothesys that all
the configurations of the fields ${\bf g}(x)$ and
$\boldsymbol{\pi}(x)$ corresponding to the same value of the
generalized action $A=\mA[{\bf g},\boldsymbol{\pi}]$ have identical
probability. This is the only and main assumption of the
paper. \ggcol{Remarkably, within the framework of symplectic
  quantization the \emph{non-renormalizability} problem of gravity is
  much more limited in scope. In particular, it does not cause
  problems to the definition of the hard-constraints ensemble for
  quantum fluctuations, Eq.~(\ref{eq:micro-gravity}).} Finally, by
considering a sort of \emph{low-energy} approximation where the
ensemble defined by a \emph{hard} constraint on the functional
$\mA[{\bf g},\boldsymbol{\pi}]$ is replaced with the ensemble
characterized by a \emph{soft} constraint on $\mA[{\bf
    g},\boldsymbol{\pi}]$ (e.g., by means of an integral transform),
we have shown that the integration over the momentum field
$\boldsymbol{\pi}(x)$ gives rise to a cosmological constant in a
theory of pure gravity initially free from it. We have therefore shown
that within the framework of symplectic quantization the appearance of
a cosmological constant $\Lambda$ in Einstein equations is still a
quantum effect, but it solely related to the intrinsic properties of
the metric field ${\bf g}(x)$, in particular to its \emph{rate of
  change} with respect to \ggcol{intrinsic} time expressed by the
momentum field $\boldsymbol{\pi}(x)$. Quite remarkably, exploiting the
definition of the DeWitt supermetric $\mG^{\mu\nu,\alpha\beta}[{\bf
    g}(x)]$ to build the kinetic term $\mK[{\bf g},\boldsymbol{\pi}]$
of symplectic quantization, $\Lambda$ turns out to be constant across
the whole space-time manifold only in $d=4$.

Thanks to the additional \emph{\ggcol{intrinsic} time} $\tau$ variable of the
symplectic quantization approach we have thus been able to
reincorporate the most natural notion of \emph{time} as a flux of
events in a quantum theory of gravity. We also provided a \emph{new}
interpretation of the cosmological constant $\Lambda$ and hopefully
defined a consistent framework for the study of cosmological problems
where an explicit dependence on time is needed. Finally, we want to
stress that the symplectic quantization of gravity, though in some
sense closer to the canonical approach to quantum gravity (it does not
require any sort of supersymmetry or extra-dimensions), does not make
any claim on the microscopic degree of freedom of the gravitational
field and can be therefore equally well compatible with different
microscopic theories.

\begin{acknowledgments}
 I thank for useful discussions V. Astuti, M. Bonvini, S. Caracciolo,
 E. De Giuli, R. Livi, A. Riotto, S. Matarrese and A. Vulpiani. I am
 expecially grateful to R. Livi and S. Matarrese for a careful reading
 of the manuscript and helpful criticism. I acknowledge Sapienza
 university of Rome, Physics Department, for kind hospitality during
 some periods in the preparation of this manuscript. Finally, I need
 to thank F. De Felice for having being such an inspiring teacher in
 the general relativity course of the academic year 2004/2005 at the
 University of Padova.
\end{acknowledgments}

\newpage

\begin{widetext}

\section{Appendix}

\subsection{Gaussian Functional Integral}
\label{App:A1}

The result of the Gaussian functional integral in Eq.~(\ref{eq:main})
can be obtained as follows. The first step is to introduce the
dimensionless integration variable
\begin{align}
\hat{\pi}_{\mu\nu}(x) =  \sqrt{\kappa_g}~\pi_{\mu\nu}(x)
\end{align}
so that the functional integral to be calculated reads as
\begin{align}
  \mZ(1) = \int \prod_x \prod_{\mu\nu} d\hat{\pi}_{\mu\nu}(x)~\exp\left( - \frac{1}{2\kappa_g^2} \int d^4x~|g|^{1/2}~\hat{\pi}_{\mu\nu}(x)~\mG^{\mu\nu,\alpha\beta}[g(x)]~\hat{\pi}_{\alpha\beta}(x) \right),
\end{align}
where, for convenience, we have considered the Euclidean version of
the path integral. It is then natural to introduce the invariant and
dimensionless infinitesimal volume element
\begin{align}
d\mu(x) = \frac{d^4x~|g|^{1/2}}{\kappa_g^2},
\end{align}  
and simply write
\begin{align}
  \mZ(1) &=
  \int \prod_x \prod_{\mu\nu} d\hat{\pi}_{\mu\nu}(x)~\exp\left( - \frac{1}{2} \int d\mu(x)~\hat{\pi}_{\mu\nu}(x)~\mG^{\mu\nu,\alpha\beta}[g(x)]~\hat{\pi}_{\alpha\beta}(x) \right) \nonumber \\
  & = \int \prod_{\bf r} \prod_{\mu\nu} d\hat{\pi}_{\mu\nu}({\bf r})~
  \exp\left( -\frac{1}{2} \sum_{\bf r}~\hat{\pi}_{\mu\nu}({\bf r})~\mG^{\mu\nu,\alpha\beta}[g({\bf r})]~\hat{\pi}_{\alpha\beta}({\bf r}) \right) \nonumber \\ 
  & = \prod_{\bf r} \frac{1}{\sqrt{\text{det}[\mG({\bf r})]}} = \exp\left( -\frac{1}{2} \sum_{\bf r} \log\text{det}[\hat{\mG}({\bf r})] \right) \nonumber \\
  & = \exp\left( -\frac{1}{2\kappa_g^2} \int d^4x~|g|^{1/2}~\log\text{det}[\hat{\mG}(x)] \right).
  \label{eq:steps-Gaussian}
\end{align}
The only subtelty of the calculation in Eq.~(\ref{eq:steps-Gaussian})
above is that we have discretized the path integral on a disordered
lattice (a lattice consistent with local Lorenz invariance) such that
each point ${\bf r}$ is surrounded by a unitary invariant volume. The
original continuum measure is restored in the last line, at the end of
the calculation, so that the particular choice of the
discretization should be irrelevant.

\end{widetext}
  

\end{document}